# Near-field photon Nernst effect

Alireza Kalantari Dehaghi[1], Linxiao Zhu[1*]

[1]Department of Mechanical Engineering, The Pennsylvania State University, University Park, PA 16802, United States

[*]Corresponding author. Email: lqz5242@psu.edu

## Abstract

We consider the consequence of having nonreciprocal photon transfer between two surfaces with temperature gradient. We demonstrate that in a system consisting of graphene and a magneto-optical substrate separated by a gap, a transverse electric field is generated in graphene perpendicular to magnetic field and temperature gradient, in analogy to the Nernst effect. Such photon Nernst effect is driven by nonreciprocal Casimir force carried by photons. We show that the thermal efficiency of near-field photon Nernst effect is bounded by the Carnot limit, and nonreciprocal high-k modes are needed for approaching the limit. The near-field photon Nernst effect enables electrical probing of nonreciprocal radiative transfer and provides a novel means of energy harvesting.

The Nernst effect, producing a transverse voltage perpendicular to both magnetic field and temperature gradient, is important for probing electronic structure [1] and energy harvesting [2]. In the Nernst effect, charges are deflected by Lorentz force. From magneto-optical materials in magnetic field, the emission of nonreciprocal thermal photons can carry asymmetric momentum [3-7]. Inspired by the importance of the Nernst effect in understanding charge transport and energy conversion, in this Letter we consider near-field photon Nernst effect. The main results of our Letter are schematically shown in Fig. 1, where we consider a system consisting of graphene as body 1 at temperature $T_1$, and a magneto-optical substrate as body 2 at temperature $T_2 \neq T_1$, separated by a vacuum gap $d$, with magnetic field along $-y$ direction. We show that with nonreciprocal Casimir force carried by photons, a transverse electric field is generated in graphene perpendicular to magnetic field and temperature gradient, as illustrated in Fig. 1(b). This contrasts with reciprocal systems without magnetic field, where there is no transverse electric field in open circuit, as shown in Fig. 1(a).

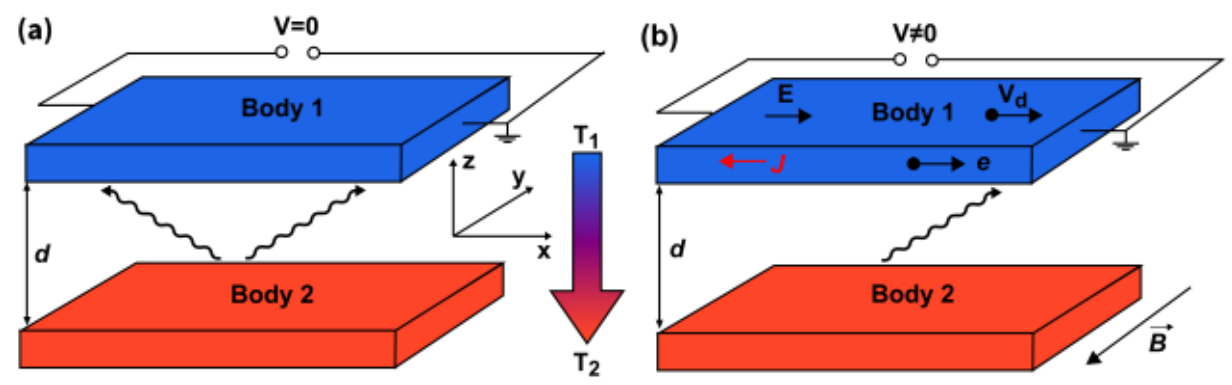

FIG. 1. Schematic of two parallel bodies 1 and 2 separated by a vacuum gap $d$ to demonstrate the near-field photon Nernst effect under orthogonal magnetic field B and temperature gradient. (a) When there is no magnetic field, there is no transverse electric field in open circuit in body 1. (b) With magnetic field $\vec{B} = -B\hat{y}$, nonreciprocal momentum transfer from body 2 to body 1 drives charge carriers in body 1 to a drift velocity $V_d$. We consider graphene as body 1, $T_1 = 5\ K$, and $T_2 = 300\ K$.

Near-field radiative transfer has attracted substantial interest [8,9]. For objects separated by a gap much smaller than thermal wavelengths (~10 μm at 300 K), the radiative heat transfer can greatly exceed the blackbody limit [8,10-23], and be quasi-monochromatic by exciting surface polaritons [24]. Accordingly, near-field radiative heat transfer shows great potential for applications such as thermophotovoltaics [25-28], photonic refrigeration [29-32], thermal diodes [33-35], and nanolithography [36].

Recently, radiative transfer involving nonreciprocal media has provided novel control of heat flow [37,38]. A range of new thermal phenomena have been introduced in magneto-optical materials or magnetic materials, such as violation of Kirchhoff's Law of thermal radiation [3-5], persistent heat current [39], photon thermal Hall effect [40], thermal routing [41], nonreciprocal near-field radiative heat transfer [42,43], and nonreciprocal thermal rectification [44]. Photons carry both energy and momentum. For nonreciprocal systems in thermal nonequilibrium with the surroundings, Casimir torque or force can arise, and accordingly, nonreciprocal systems have been proposed to function as heat engines converting heat to mechanical work [45-47]. A thermoelectric effect was indicated between two plates in magnetic field [6], but without analysis. Therefore, it is unclear if heat can be converted to electricity using nonreciprocal photon transfer.

We start by introducing the necessary conditions for achieving photon Nernst effect. First, the system needs to support nonreciprocal photon transfer to have a net lateral force. For systems involving two planar surfaces satisfying Lorentz reciprocity, the lateral Casimir force has mirror symmetry, leading to zero net lateral force. Nonreciprocal systems can be constructed by using magneto-optical materials in magnetic field, or using graphene with drift current [48]. For magneto-optical materials, we consider n-InSb, due to its low effective mass and high mobility. Second, the charge mobility needs to be high, such that the lateral Casimir force can drive charges to a high drift velocity leading to high current. We consider graphene owing to its high mobility [49,50].

Using fluctuational electrodynamics [51,52], the lateral Casimir force on graphene per unit area [47,53] is

$$F = \int_0^\infty \frac{d\omega}{2\pi} \int_{|k_\parallel|>\omega/c}^{\infty} \frac{d^2 k_\parallel}{4\pi^2} f_x(\omega, k_\parallel, d), \tag{1}$$

where the energy and momentum-resolved force $f_x(\omega, k_\parallel, d) = \hbar k_x n_{21}(\omega, k_x)\phi_{21}(\omega, k_\parallel, d)$, and net heat transfer from the magneto-optical substrate to graphene per unit area is

$$Q = \int_0^\infty \frac{d\omega}{2\pi} \int_{|k_\parallel|>\omega/c}^{\infty} \frac{d^2 k_\parallel}{4\pi^2} q(\omega, k_\parallel, d), \tag{2}$$

where the energy and momentum-resolved heat transfer $q(\omega, k_\parallel, d) = \hbar\omega n_{21}(\omega, k_x)\phi_{21}(\omega, k_\parallel, d)$. Here, $\omega$ is the angular frequency, $k_\parallel = (k_x, k_y)$ is the in-plane wavevector, and the difference in photon occupation numbers between body 2 and body 1 is

$$n_{21}(\omega, k_x) = \frac{1}{e^{\frac{\hbar\omega}{k_B T_2}} - 1} - \frac{1}{e^{\frac{\hbar(\omega - V_d k_x)}{k_B T_1}} - 1}, \tag{3}$$

where $V_d$ is the drift velocity of charges in graphene, and $V_d \leq V_F$ where Fermi velocity $V_F = 10^6\ m/s$. $V_d k_x$ acts as a chemical potential of photons for body 1 due to the Doppler effect [54,55]. The photon transmission coefficient $\phi_{21}(\omega, k_\parallel, d)$ is

$$\phi_{21}(\omega, k_\parallel, d) = \frac{4Im(\rho_1)Im(\rho_2)e^{2ik_z d}}{|1 - \rho_1\rho_2 e^{2ik_z d}|^2}, \tag{4}$$

where $\rho_i$ (i=1 or 2) is the reflection coefficient from vacuum to body $i$ in p polarization, and $k_z = \sqrt{k_0^2 - k_\parallel^2}$ is the $z$-component of wavevector in vacuum. Here, $k_0 = \omega/c$, and $c$ is the velocity of light in vacuum. As p-polarized surface waves dominate near-field energy and momentum transfer, it is sufficient to only consider p-polarized evanescent waves with $|k_\parallel| > k_0$ [6]. For evanescent waves, $\phi_{21}(\omega, k_\parallel, d)$ describes the photon transmission coefficient between body 2 and the finite-thickness graphene (see proof in the Supplemental Material [56]). When performing the momentum-space integral in Eqs. (1) and (2), we integrate over Cartesian coordinates $k_x$ and $k_y$, so as to ensure that the integration over $k_x$ including the contribution at $k_x \approx \omega/V_d$ has converged (see behavior of $f_x$ at large wavevectors in [56]).

The dielectric function of graphene is modeled as $\epsilon_1 = 1 + \frac{i\sigma_g}{\epsilon_0\omega\delta}$ [6,59-61]. The conductivity of graphene $\sigma_g(\omega, k_\parallel, \tau_g, V_d)$ depends on electron relaxation time $\tau_g$ and $V_d$, and is provided in Appendix. The p-polarized reflection coefficient from vacuum to the finite-thickness graphene is $\rho_1 = r_1\frac{1-e^{2ik_{z1}\delta}}{1-r_1^2 e^{2ik_{z1}\delta}}$ (see derivation in [56]), where $r_1 = \frac{\epsilon_1 k_z - k_{z1}}{\epsilon_1 k_z + k_{z1}}$, $k_{z1} = \sqrt{\epsilon_1 k_0^2 - k_\parallel^2}$, and the thickness of graphene $\delta$ is $0.3\ nm$. We consider electron doping concentration of $2 \times 10^{15}\ m^{-2}$ for the graphene, with Fermi energy $E_F = 52.2\ meV$. Due to Pauli blocking and for the small drift velocity in this work ($V_d \ll V_F$), at energy below $2E_F$, optical absorption in graphene is dominated by intraband transitions of free electrons. Further, photon transfer at >100 meV is verified to contribute to less than 0.4% for the heat transfer (see details in [56]). Accordingly, we consider photons up to 100 meV for calculating force and heat transfer using Eqs. (1) and (2).

Therefore, the lateral Casimir force and heat transfer in this work are due to intraband transitions, and the lateral Casimir force $F$ acts on the free electrons of graphene. The permittivity tensor of n-InSb is

$$\overleftrightarrow{\epsilon}(\omega) = \begin{bmatrix} \epsilon_d & 0 & i\epsilon_a \\ 0 & \epsilon_p & 0 \\ -i\epsilon_a & 0 & \epsilon_d \end{bmatrix}, \tag{5}$$

where $\epsilon_d$, $\epsilon_a$, and $\epsilon_p$ are provided in Appendix. We follow Ref. [62] to evaluate the p-polarized reflection coefficient $\rho_2$ from vacuum to the magneto-optical substrate.

When the two bodies are separated by a nanoscale gap, the momentum and heat transfer between them are dominated by gap surface polariton. The dispersion relation of the gap surface polariton at $k_y = 0$ is (see derivation in [56])

$$1 + \sigma_g \frac{k_z}{\omega\epsilon_0} + \frac{\left(1 - e^{2ik_zd}\right)(\epsilon_d k_{z2} - i\epsilon_a k_x) + \left(1 + e^{2ik_zd}\right)(\epsilon_d^2 - \epsilon_a^2)k_z}{(1 + e^{2ik_zd})(\epsilon_d k_{z2} - i\epsilon_a k_x) + (1 - e^{2ik_zd})(\epsilon_d^2 - \epsilon_a^2)k_z} = 0, \tag{6}$$

where $k_{z2} = \sqrt{\left(\epsilon_d - \frac{\epsilon_a^2}{\epsilon_d}\right)k_0^2 - k_x^2}$. Without magnetic field, the dispersion relation of the gap surface polariton at 1 nm gap is symmetric between $k_x$ and $-k_x$ (Fig. 2(a)), leading to symmetric energy transfer. However, as photons at $k_x$ and $-k_x$ carry opposite transverse momentum, their contribution to lateral force is opposite (Fig. 2(b)). Figure 2(c) shows that the photon transmission coefficient is isotropic in $(k_x, k_y)$ plane at energy 19 $meV$ corresponding to a mode (Fig. 2(a-b)). Accordingly, the spectral radiative heat transfer shows resonance peaks, but the lateral Casimir force is zero (Fig. 2(d)).

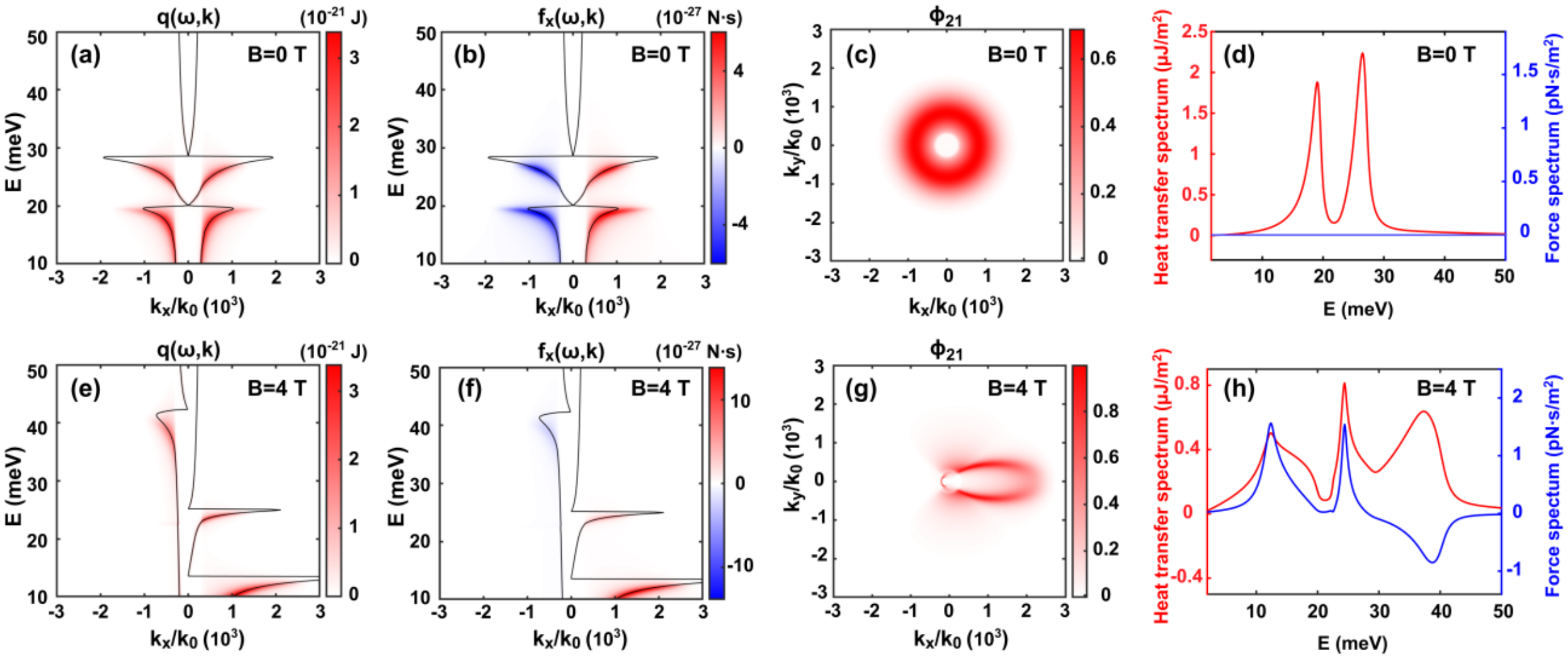

FIG. 2. Dispersion relation, and spectra of heat transfer and Casimir force. (a) Energy, momentum-resolved heat transfer and (b) lateral Casimir force, at zero magnetic field and $k_y = 0$. (c) Photon transmission coefficient at 19 $meV$. (d) Heat transfer and Casimir force spectra. In (a-d), $B = 0\ T$. (e) Energy, momentum-resolved heat transfer and (f) Casimir force, at 4 $T$ and $k_y = 0$. (g) Photon transmission coefficient at 12.4 $meV$. (h) Heat transfer and Casimir force spectra. In (e-h), $B = 4\ T$. (a,e) include gap polariton dispersion relation as black lines. In (a-h), the gap is 1 nm, and graphene has zero drift velocity and $2 \times 10^{15}\ m^{-2}$ electron concentration.

In contrast, magnetic field breaks the mirror symmetry of the dispersion relation. In magnetic field, the peak of zero-field heat transfer spectrum at 19 meV splits to two branches, and one branch red shifts to 12.4 meV at 4 T (Fig. 2(e); see details in [56]). Accordingly, there is strong radiative heat transfer and lateral force at positive wavevector in Fig. 2(e-f), but weak contribution from negative wavevector. Figure 2(g) further shows that the photon transmission coefficient at 12.4 $meV$ is highly asymmetric. Accordingly, lateral force is non-zero (Fig. 2(h)), showing two positive peaks at 12.4 $meV$ and 24.4 $meV$, and a negative peak at 38.9 $meV$ (Fig. 2(f)). Therefore, asymmetric dispersion relation in nonreciprocal system is crucial to having nonzero lateral Casimir force.

A positive lateral Casimir force will drive free electrons in graphene towards positive $x$, leading to accumulation of negative charges in positive $x$ and subsequently establishment of an internal electric field $E$ towards the $+x$ direction. For free electrons of graphene in force equilibrium, we have

$$F - n_g eE - \frac{n_g e V_d}{\mu_g} = 0, \tag{7}$$

where $-n_g eE$ is the electric force on free electrons per unit area [54], $n_g$ is electron doping concentration of graphene, and $e$ is the charge of a proton. Free electrons of graphene at drift velocity $V_d$ are subject to friction-related force per unit area $-\frac{n_g e V_d}{\mu_g}$ [55] (see derivation in [56]), where $\mu_g$ is the charge mobility. Such friction-related force on free electrons is due to their collision with other particles, including lattice vibrations. By connecting transverse ends of graphene to an external load, the system can produce electrical work. The electrical power generation rate per unit area is $P = -EJ$, where current density $J = -n_g e V_d$. The thermal efficiency is $\eta = P/Q$.

We next show that efficiency is upper bounded by the Carnot limit. Using Eq. (7), expressions of $P$ and $\eta$, we have

$$\eta = \frac{-V_d^2 \frac{n_g e}{\mu_g} + \int_0^\infty \frac{d\omega}{2\pi} \int \frac{d^2 k_\parallel}{4\pi^2} \hbar k_x V_d n_{21} \phi_{21}}{\int_0^\infty \frac{d\omega}{2\pi} \int \frac{d^2 k_\parallel}{4\pi^2} \hbar \omega n_{21} \phi_{21}}. \tag{8}$$

As Joule heating term $V_d^2 \frac{n_g e}{\mu_g}$ is non-negative and heat transfer $Q > 0$, we have

$$\eta \leq \frac{\int_0^\infty \frac{d\omega}{2\pi} \int \frac{d^2 k_\parallel}{4\pi^2} \hbar k_x V_d n_{21} \phi_{21}}{\int_0^\infty \frac{d\omega}{2\pi} \int \frac{d^2 k_\parallel}{4\pi^2} \hbar \omega n_{21} \phi_{21}}. \tag{9}$$

It follows that

$$\frac{T_2}{T_1}(1-\eta) - 1 \geq \frac{k_B T_2 \int_0^\infty \frac{d\omega}{2\pi} \int \frac{d^2 k_\parallel}{4\pi^2} \left[\frac{\hbar(\omega - V_d k_x)}{k_B T_1} - \frac{\hbar\omega}{k_B T_2}\right] \left[\frac{1}{e^{\frac{\hbar\omega}{k_B T_2}} - 1} - \frac{1}{e^{\frac{\hbar(\omega - V_d k_x)}{k_B T_1}} - 1}\right] \phi_{21}}{Q}. \tag{10}$$

For wavevector $k_x < \omega/V_d$, graphene is dissipative with $Re(\sigma_g) > 0$ and $\phi_{21} > 0$, leading to $\left[\frac{\hbar(\omega - V_d k_x)}{k_B T_1} - \frac{\hbar\omega}{k_B T_2}\right] \left[\frac{1}{e^{\frac{\hbar\omega}{k_B T_2}} - 1} - \frac{1}{e^{\frac{\hbar(\omega - V_d k_x)}{k_B T_1}} - 1}\right] \phi_{21} > 0$. When $k_x > \omega/V_d$, for graphene with small damping $\gamma_g \to 0$, negative Landau damping appears, leading to $Re(\sigma_g) < 0$ [6,63], negative photon transmission coefficient $\phi_{21} < 0$, and accordingly $\left[\frac{\hbar(\omega - V_d k_x)}{k_B T_1} - \frac{\hbar\omega}{k_B T_2}\right]\left[\frac{1}{e^{\frac{\hbar\omega}{k_B T_2}}-1} - \frac{1}{e^{\frac{\hbar(\omega - V_d k_x)}{k_B T_1}}-1}\right]\phi_{21} > 0$. Therefore, the integrand in the numerator of Eq. (10) is always non-negative, and Eq. (10) gives $\frac{T_2}{T_1}(1-\eta) - 1 \geq 0$, which leads to $\eta \leq \eta_{Carnot}$ where $\eta_{Carnot} = 1 - T_1/T_2$. Therefore, the thermal efficiency of near-field photon Nernst effect is always bounded by the Carnot limit.

From Eqs. (8) and (10), it can be seen that the Carnot efficiency can be approached when the Joule heating is negligible which requires high charge mobility of graphene, and also the photon transmission coefficient $\phi_{21}$ satisfies the following dispersion relation

$$\frac{\hbar(\omega - V_d k_x)}{k_B T_1} = \frac{\hbar\omega}{k_B T_2}. \tag{11}$$

Equation (11) can be rewritten as $k_x = \frac{\omega}{V_d}\eta_{Carnot} = k_0 \cdot \frac{c}{V_F} \cdot \frac{V_F}{V_d}\eta_{Carnot} \approx 300 k_0 \cdot \frac{V_F}{V_d}\eta_{Carnot}$. As $V_F \geq V_d$, we have $k_x \geq 300 k_0 \cdot \eta_{Carnot}$. When $\eta_{Carnot} > \frac{1}{300}$ which applies for significant temperature difference, the optimal dispersion relation requires $k_x$ to exceed $k_0$. Such transverse momentum exceeding $k_0$ cannot be supported by propagating photons in free space. In contrast, in the near field, high $k_x \gg k_0$ can be supported in surface modes. Therefore, when $\eta_{Carnot} > \frac{1}{300}$, near-field radiative transfer is crucial to approaching the Carnot efficiency and to achieving strong photon Nernst effect.

Based on the principles introduced above, we numerically assess electrical power generation using near-field photon Nernst effect. Figure 3(a) shows the relation between current density and electric field, for 1 nm gap, and $2 \times 10^{15}\ m^{-2}$ electron doping concentration in graphene, corresponding to an electron mobility of 23 $m^2V^{-1}s^{-1}$ [49]. The current versus electric field curves show a positive slope, which can be understood from the dependence of lateral Casimir force on the drift velocity, and electrical resistance of graphene. When $V_d$ increases with decreasing $J$, the chemical potential of photons for body 1 ($V_d k_x$) increases for positive $k_x$ and decreases for negative $k_x$, leading to reduced $\hbar k_x n_{21}$ regardless of the sign of $k_x$, which reduces the Casimir force $F$. Using $E = \frac{F}{n_g e} - \frac{V_d}{\mu_g}$ from Eq. (7), the combination of increasing $V_d$ and decreasing $F$ will reduce $E$, leading to the positive slope of J versus E.

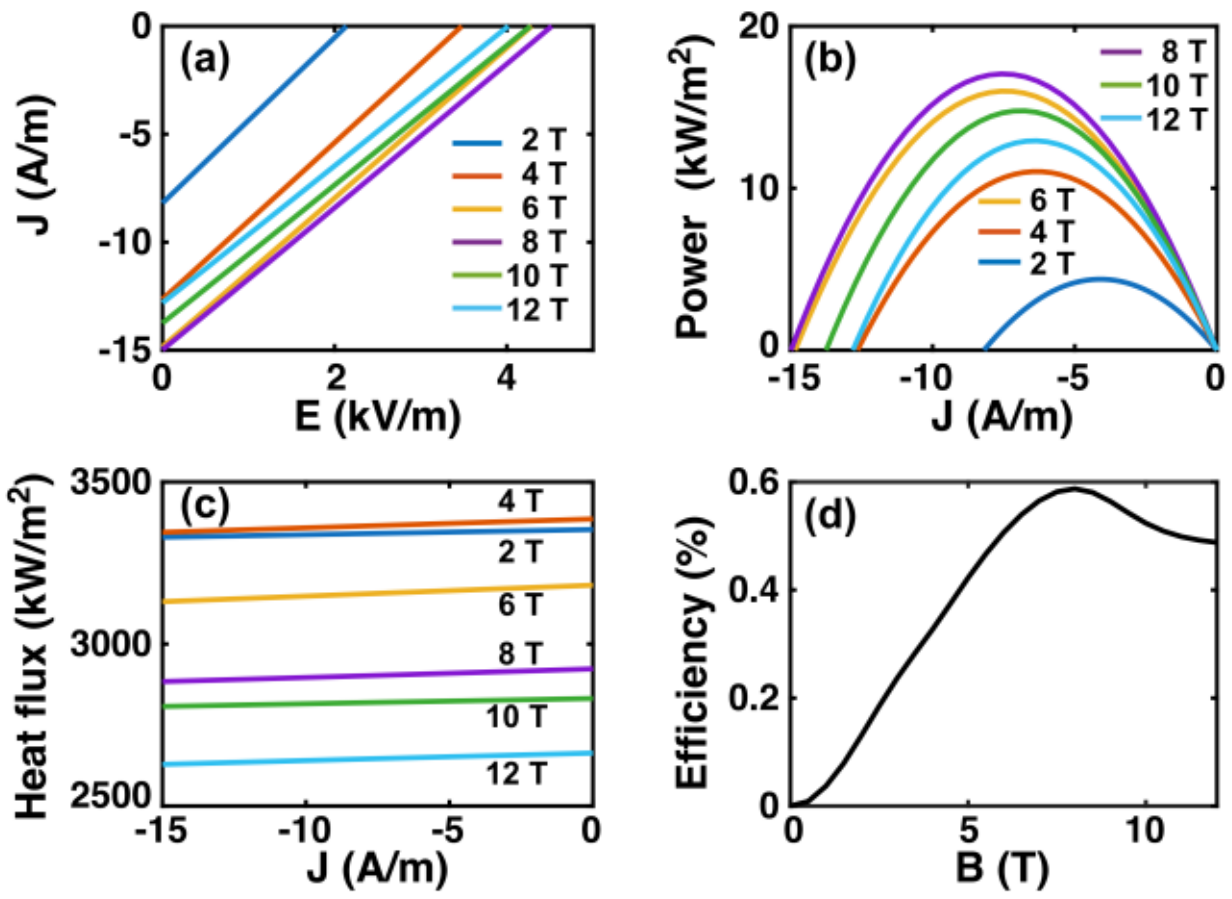

FIG. 3. Magnetic-field dependence of J-E curves, power generation, heat transfer, and efficiency. (a) Relation between current density $J$ and electric field $E$ at different magnetic fields. (b) Power generation rate and (c) radiative heat transfer rate as a function of $J$ at different magnetic fields. (d) Optimal power conversion efficiency as a function of magnetic field. In (a-d), the gap is 1 nm, and the electron doping concentration of graphene is $2 \times 10^{15}\ m^{-2}$.

We observe that the open-circuit electric field is non-zero with external magnetic field and is zero without magnetic field (Fig. 3(a); see relation between open-circuit electric field and magnetic field in [56]). Such generation of transverse open-circuit electric field orthogonal to magnetic field and temperature gradient is directly analogous to the Nernst effect.

When the magnetic field $B \leq 8\ T$, the power generation rate increases with the magnetic field (Fig. 3(b)), due to increasing electric field (Fig. 3(a)). At larger magnetic field, however, the power generation rate decreases. Figure 3(c) shows that the radiative heat transfer decreases with the increasing magnitude of the current. When the magnetic field exceeds 4 $T$, the heat transfer decreases as the magnetic field increases. Figure 3(d) shows the optimal efficiency as a function of magnetic field at 1 nm gap. The efficiency has non-monotonic dependence on magnetic field, and it reaches the maximum of 0.59% at 8 $T$. At small magnetic field ≤8 T, as the magnetic field increases, the power generation increases and the heat transfer decreases (Fig. 3(b-c)), leading to increasing efficiency. At high magnetic fields, as the magnetic field increases, the power generation substantially decreases and the heat transfer is relatively stable, leading to reduced efficiency.

As photon Nernst effect is most efficient in the near field, a key question in this context is how the effect depends on the gap. To answer this we study the dependence of the near-field photon Nernst effect on the gap. Figure 4(a) shows the dependence of efficiency on the gap. As the gap reduces from 100 nm to 1 nm, the efficiency increases, reaching the maximum of 0.59% at $1\ nm$ gap and $8\ T$. The gap dependence of efficiency results from the different dependence of power generation and heat transfer on the gap. As the gap reduces, the radiative heat transfer rate increases (Fig. 4(c)) due to the contribution of high-k modes in the near field. The power generation shows steeper gap dependence than heat transfer (Fig. 4(b)), which can be explained by considering gap dependence of the open-circuit electric field and weak dependence of J-E slope on gap. First, because the open-circuit electric field is proportional to the open-circuit Casimir force (Eq. (7)), whose integrand scales as the photon transmission coefficient (Eq. (1)), the open-circuit electric field has similar gap dependence as the heat flux. Further, the J-E curve is approximately linear (Fig. 3(a)), and the slope of J-E curves as shown in Fig. 4(d) only weakly depends on gap. The J-E slope reduces from $\mu_g n_g e = 0.0074\ \Omega^{-1}$ at gap >10 nm to $\sim 0.004\ \Omega^{-1}$ at 1 nm. Therefore, the optimal power generation rate scales approximately quadratically with the open-circuit electric field, yielding a stronger near-field enhancement than the heat flux, which scales approximately linearly with the open-circuit electric field, and consequently leading to increased efficiency as the gap reduces.

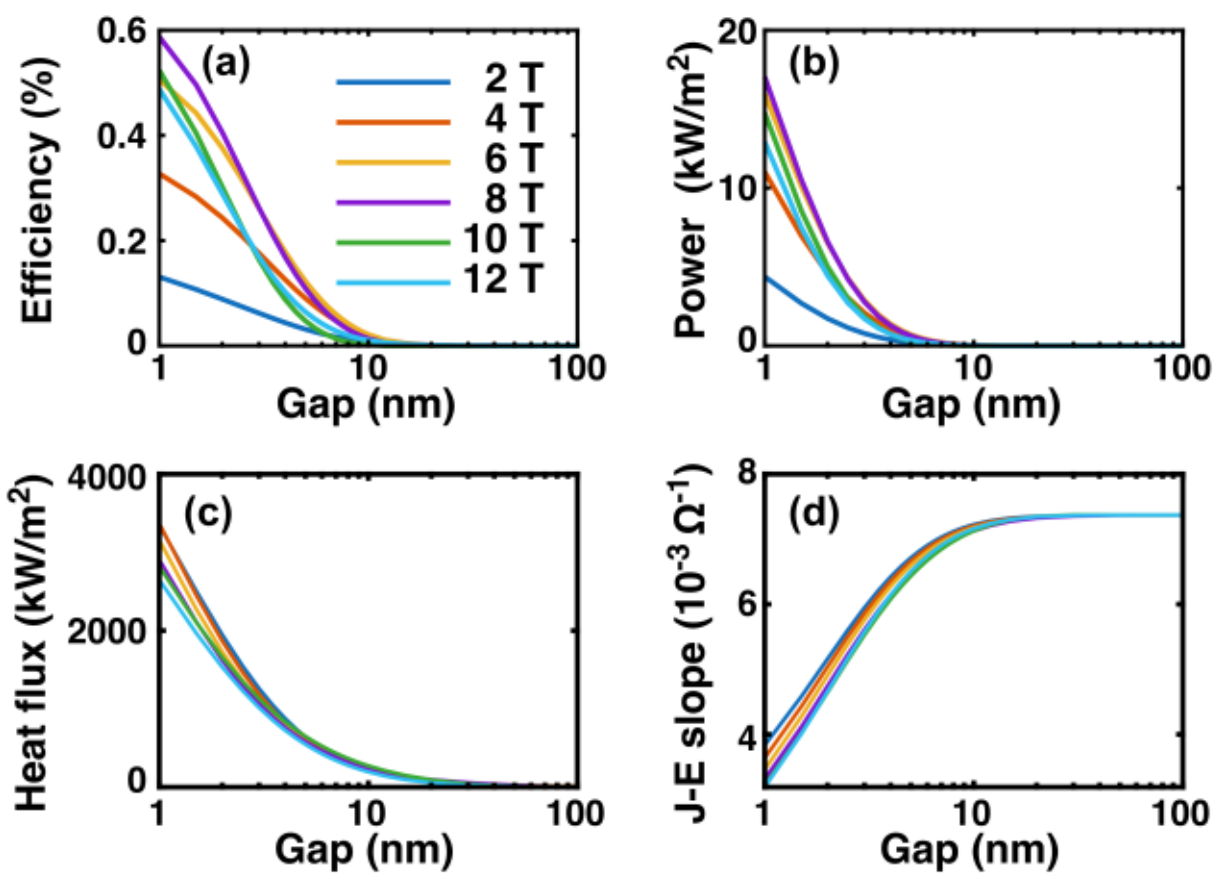

FIG. 4. Gap dependence of the near-field photon Nernst effect. (a) Efficiency as a function of gap at different magnetic fields, with (b) the corresponding power generation rate and (c) radiative heat

transfer rate. (d) Slope of J-E curves as a function of gap. In (a-d), the electron doping concentration in graphene is $2 \times 10^{15}\ m^{-2}$, and the legend in (a) is used throughout (a-d).

We further discuss photon Nernst effect for graphene at room temperature, and the behavior at small temperature difference. We consider graphene at 300 K, with graphene electron mobility $20\ m^2V^{-1}s^{-1}$ at the room-temperature intrinsic mobility limit [64,65] and graphene electron concentration $2 \times 10^{15}\ \mathrm{m}^{-2}$. The magneto-optical medium is at 330 K. The estimated efficiency reaches 0.01% at 1 nm gap and 8 $T$, which is 0.12% of the Carnot limit. Therefore, near-field photon Nernst effect is also expected using graphene at room temperature. We further derive the efficiency at the small temperature limit as a function of drift velocity (see details in [56]), by performing a Taylor expansion for $n_{21}$ to the first order of temperature difference and drift velocity while neglecting the dependence of the photon transmission coefficient on drift velocity. The analytical result agrees well with direct calculation (see details in [56]).

To conclude, we introduce near-field photon Nernst effect in a system consisting of graphene and a magneto-optical substrate, where a transverse electric field is generated perpendicular to both magnetic field and temperature gradient. We show that the thermal efficiency of photon Nernst effect is upper bounded by the Carnot limit and show that near-field radiative transfer is crucial to approaching the limit. To achieve the Carnot limit for the near-field photon Nernst effect, Joule heating needs to be suppressed such as by enhancing the charge mobility of graphene. Higher mobility in graphene can be achieved at lower carrier concentrations [49,50,66], which will be an intriguing regime for exploring the effect, while considering both intraband and interband transitions. While we considered suspended graphene for simplicity, for graphene supported by a substrate [67], a heat flux ~0.34 kW/cm$^2$ (Fig. 3) is estimated to result in a negligible electronic temperature rise of 6 K in graphene. Moreover, non-equilibrium phonon distribution in graphene is known to be insignificant up to a temperature rise of 400 K [67]. We note that a thermal efficiency of 0.07% is predicted at 1 $T$, which can be achieved using ferromagnets [68]. As generating non-zero open-circuit electric field requires breaking the Lorentz reciprocity, which can be achieved in magnetic Weyl semimetals intrinsically [4], magnetic Weyl semimetals may enable generating finite open-circuit electric field without any magnetic field. We considered a local permittivity for the magneto-optical medium; nonlocal corrections may become relevant at gap sizes below a few nanometers, but the prediction of the photon Nernst effect qualitatively holds.

**Acknowledgments**

L.Z. acknowledges partial support from National Science Foundation award CBET-2238927. Computations for this research were performed on the Pennsylvania State University's Institute for Computational and Data Sciences' Roar Collab supercomputer.

## End Matter

*Appendix – Conductivity of graphene:* Without collision and drift current, the zero-temperature nonlocal graphene conductivity [69,70] is

$$\sigma_g(\omega, k_\parallel) = -\frac{i\omega e^2}{4\pi\hbar}\left[\frac{8E_F}{\hbar V_F^2 k_\parallel^2} + \frac{[G(-\Delta_-) - i\pi]\theta(-\Delta_- - 1) + G(\Delta_-)\theta(\Delta_- + 1) - [G(\Delta_+) - i\pi]}{\sqrt{\omega^2 - k_\parallel^2 V_F^2}}\right]. \tag{12}$$

Here $G(z) = z\sqrt{z^2 - 1} - \ln(z + \sqrt{z^2 - 1})$, $\Delta_\pm = (\hbar\omega \pm 2E_F)/(\hbar V_F k_\parallel)$, $E_F = \hbar V_F\sqrt{\pi n_g}$ is the chemical potential [71], $n_g$ is electron concentration of graphene, and $\theta(x)$ is unit step function. After accounting for collision and a Doppler shift for electrons moving at velocity $V_d$ along $x$ direction [72], the graphene conductivity is

$$\sigma_g(\omega, k_\parallel, \tau_g, V_d) = -i\omega \frac{\left[1 + \frac{i/\tau_g}{\omega - k_x V_d}\right] \chi(\omega - k_x V_d + i/\tau_g, k_\parallel)}{1 + \frac{i/\tau_g}{\omega - k_x V_d} \chi(\omega - k_x V_d + i/\tau_g, k_\parallel)/\chi(0, k_\parallel)}, \tag{13}$$

where $\chi(\omega, k_\parallel) = \sigma_g(\omega, k_\parallel)/(-i\omega)$ is susceptibility, and relaxation time $\tau_g = \hbar\mu_g\sqrt{\pi n_g}/(eV_F)$ [71]. In this work, we use Eq. (13) to describe the graphene conductivity.

*Permittivity components of the magneto-optical medium*: The permittivity components of n-InSb in Eq. (5) are

$$\epsilon_d = \epsilon_{ib} + \epsilon_\infty \left[1 + \frac{\omega_L^2 - \omega_T^2}{\omega_T^2 - \omega^2 - i\Gamma\omega} + \frac{\omega_p^2(\omega + i\gamma_{MO})}{\omega[\omega_c^2 - (\omega + i\gamma_{MO})^2]}\right], \tag{14}$$

$$\epsilon_a = \frac{-\epsilon_\infty \omega_c \omega_p^2}{\omega[(\omega + i\gamma_{MO})^2 - \omega_c^2]}, \tag{15}$$

$$\epsilon_p = \epsilon_{ib} + \epsilon_\infty \left[1 + \frac{\omega_L^2 - \omega_T^2}{\omega_T^2 - \omega^2 - i\Gamma\omega} - \frac{\omega_p^2}{\omega(\omega + i\gamma_{MO})}\right]. \tag{16}$$

Here, $\epsilon_\infty = 15.7$, $\omega_L = 3.62 \times 10^{13}\ rad/s$, $\omega_T = 3.39 \times 10^{13}\ rad/s$, $\Gamma = 5.65 \times 10^{11}\ rad/s$, and $\epsilon_{ib}$ is the interband contribution at energy above the bandgap [73]. For the magneto-optical medium, we consider an electron concentration $n_{MO} = 1.42 \times 10^{17}\ cm^{-3}$, electron effective mass $m_{MO}^* = 0.018\ m_e$ where $m_e$ is electron mass [74], electron mobility $\mu_{MO} = 4.48\ m^2/(V.s)$ [49] leading to electron damping rate $\gamma_{MO} = \frac{e}{m_{MO}^* \mu_{MO}} = 2.18 \times 10^{12}\ rad/s$, the cyclotron frequency $\omega_c = \frac{eB}{m_{MO}^*}$, and the plasma frequency $\omega_p = \sqrt{\frac{n_{MO} e^2}{\epsilon_\infty m_{MO}^*}} = 4.00 \times 10^{13}\ rad/s$.